\documentclass[useAMS,usenatbib]{pasa}
\usepackage{graphicx}
\title[Cosmic Biology]{Cosmic Biology in Perspective}
\bibpunct{(}{)}{;}{a}{}{,}

\author[Wickramasinghe et al.]{N. C. Wickramasinghe$^{1,6}$, Dayal T. Wickramasinghe$^2$, Christopher A. Tout$^{3,4}$, John C. Lattanzio$^{4}$
 and Edward J. Steele$^{5}$
\affil{$^1$Buckingham Centre for Astrobiology, University of Buckingham, UK}
\affil{$^2$Mathematical Sciences Institute, The Australian National University, ACT 0200, Australia}
\affil{$^3$Institute of Astronomy, The Observatories, Madingley Road, Cambridge CB3 OHA}
\affil{$^4$Monash Centre for Astrophysics, School of Physics and Astronomy, Monash University, 10 College Walk, Clayton, VIC 3168, Australia}
\affil{$^5$C Y O'Connor Village Foundation, 4D, 11A Erade Drive, Piara Waters, WA 6112, Australia}
\affil{$^6$Centre for Astrobiology, University of Ruhuna, Matara, Sri Lanka}
}

\jid{PASA}
\doi{10.1017/pas.\the\year.xxx}
\jyear{\the\year}

\usepackage{aas_macros}
\def\Msun{\ifmmode{\,M_\odot}\else$\,M_\odot$\fi}
\def\msun{\ifmmode{\,M_\odot}\else$\,M_\odot$\fi}

\begin{document}

\begin{frontmatter}
\maketitle

\begin{abstract}
A series of astronomical observations obtained over the period 1986 to
2018 supports the idea that life is a cosmic
rather than a purely terrestrial or planetary phenomenon. These include
(1)~the detection of biologically relevant molecules in interstellar
clouds and in comets, (2)~mid-infrared spectra of interstellar grains
and the dust from comets, (3) a diverse set of data from comets
including the {\it Rosetta} mission showing consistency with biology
and (4)~the frequency of Earth-like or habitable planets in the
Galaxy.  We argue that the conjunction of all the available data
suggests the operation of cometary biology and interstellar panspermia
rather than the much weaker hypothesis of comets being only the source of
the chemical building blocks of life.  We conclude with specific
predictions on the properties expected of extra-terrestrial life if it
is discovered on Enceladus, Europa or beyond.  A radically different
biochemistry elsewhere can be considered as a falsification of the
theory of interstellar panspermia.
\end{abstract}

\end{frontmatter}

\section{INTRODUCTION }
\label{sec:intro}

The essence of biological evolution confined to the Earth is that the
accumulation of random copying errors in a genome leads, after the
lapse of very many generations, to the emergence of new phenotypes,
new species, new orders, new classes.  It is postulated that, in the
process of replication, a random mutation, that is favoured for
survival, occurs and is locked into the genome.  Subsequently another
random mutation is similarly selected for survival and
locked into the germ-line, and the process repeats with no external
input required.  However, the failure thus far to synthesise living
cells nor indeed any approach to a living system de novo in the
laboratory after over five decades of serious effort is discouraging
\citep {Deamer2012}.

The sufficiency of this neo-Darwinian scheme operating in isolation
may be called into question by the observation that, following the
first appearance of bacterial life (prokaryotes) on the Earth between
4.1 and~4.2\,Gyr ago \citep{Bell2015}, there is a protracted
geological period of stasis lasting for about~2\,Gyr during which no
further evolution is evident.  During this period, bacteria and
archaea remain the sole life forms on the Earth.  The oldest evidence,
from molecular data, of further evolution to eukaryotes is found only
2\,Gyr ago \citep{Brocks1999, Strother2011}.  Eukaryotes differ
fundamentally from bacteria and archaea.  Besides being larger in
size, eukaryotes possess nuclei and organelles, which are thought to
have been symbiotic bacteria, and they can reproduce sexually.

Following the appearance of eukaryotes on the Earth at~2\,Gyr there is
another protracted period in which they remain essentially unaltered.
This is also difficult to explain if conventional Earth-centred
evolution is the main driving force.  Some attempts at multicellular
cooperation appears in the form of sheets or filaments but these
experiments appear to have come to a dead end \citep{Knoll2011}. The
next major innovation in terrestrial biology is the emergence of
multicellular life, small animals, and this comes only after another
long period of stasis lasting for some 1.5\,Gyr until the Cambrian
explosion about 540\,Myr ago.  Again we can note that 1.5\,Gyr of
neo-Darwinian evolution showed no signs of any progress towards
multi-cellularity.  Subsequent progress in biological evolution also
seems to be punctuated by a series of sharp spikes of speciation as
well as of extinctions extending arguably to recent geological epochs
\citep{Steele2018}. The current explanation for this is that mass
extinction opens up many opportunities for new life.

It has been estimated that the solar system has experienced dynamical
encounters with 5 to~10 giant molecular clouds throughout the course
of its history, passing within~5\,pc of star-forming nebulae on the
average about once every~100\,Myr.  Numerical simulations with
standard $N$-body techniques have led to the conclusion that the
ingress of comets into the inner solar system is increased by factors
of about~100 during each such encounter \citep{Wick2009}.  Such
enhanced collision rates with comets have the potential to introduce
non-solar system biological genetic material.  The neo-Darwinian
scheme, augmented by horizontal gene transfer during such events, may
lead to a general accord with patterns of biological evolution that
are witnessed on the Earth \citep{WallisWick2004}.  The same process
can be extended to take place on a Galactic scale between the 10
thousand million or so habitable planetary systems that are now
thought to exist in our Galaxy alone \citep{Kopparapu2013}.

Within the solar system itself, and on the Earth in particular, the
evolution of life would be controlled by a combination of processes,
random encounters with dense star/planet forming clouds in the course
of its motion through the Galaxy, and the more or less periodic
incursions of biologically laden material from comets within the solar
solar system itself.  In this way it may be possible to account for
both random events, involving long periods of stasis such as we have
discussed, as well as periodic events that punctuate the evolution of
life.

\section{Interstellar Carbon}
\label{sec:int carbon}


It is now accepted that interstellar dust includes a component of
organic grains \citep{Wick1974}.  In particular, organic grains
constitute an integral part of progenitor clouds, similar to the
presolar nebula, from which the Sun and planets formed 4.8\,Gyr
ago. Icy cometesimals are the first condensed bodies that formed in
the outer regions of such a nebula and these bodies can be assumed to
have mopped up a large fraction of interstellar organics.  Such
organic molecules would then have been available for delivery via
collisions of comets to the inner planets of the solar system
including Earth.

\subsection{Organic molecules in space}

Beyond the attribution of hydrocarbon structures in the infrared
spectra of Galactic sources to interstellar dust (section~3.1),
the definite detections of specific organic molecular species by
millimetre wave and radio observations have revealed well over 150
such molecules \citep{Kwok2016, Thaddeus2006}.  They include a branched
carbon back-bone isopropyl cyanide, the amino acid glycine, vinyl
alcohol, ethyl formate, naphthalene and the chiral molecule propylene
oxide \citep{Kuan2003}.

From the 1930s onwards observed spectra of reddened stars have shown a
large number of unidentified diffuse interstellar bands (DIBs)
extending over visual and near IR wavelengths \citep{Herbig1995}.
Complex organic molecules, including even molecules related to
chlorophyll, have been proposed but conclusive identification of a
specific carrier has remained elusive \citep{Johnson2006}.  Two
spectral lines at $9577\,${\AA} and $9632\,${\AA} match a singly
charged fullerine $\rm{C_{60}^{+}}$ \citep{Campbell2015}, although
such an identification of only~2 out of some~400 similar lines can be
questioned.  More recently there have been claims for the presence of
other features attributed to fulleranes in the spectrum of the
proto-PNe IRAS01005+7910 \citep{Zhang2016}.  Fullerine cages that have
also been discovered in carbonaceous chondrites provide an additional
line of evidence.

The organic molecules so far discovered in interstellar space have
prompted astrobiologists to consider seriously the proposition that
they may somehow be connected with the origin of terrestrial life.
However, even if the appropriate organic molecules are supplied to
Earth's oceans where they can contribute to a canonical primordial
soup, there remains the improbability of small-scale abiogenesis
\citep{HW81}.  For this reason, \cite{Crick1973} advocated a directed
panspermia where life first originated somewhere else in the larger
Universe.

\subsection{The origin of interstellar grains}

The question of the nature and mode of origin of interstellar grains
has been addressed since the 1930s \citep{Lyndwick68,Draine2003}.
The classic work of \cite{Oort1946} laid the foundations of modern
research on grain formation and grain mantle growth in interstellar
clouds.  However, difficulties associated with the nucleation of dust
in low-density interstellar clouds led to a serious discussion of
alternative sites of grain formation.  \cite {HW62,HW69} argued that
mass outflows from cool giant stars, in particular carbon stars,
provide ideal conditions for the nucleation and growth of refractory
grains.  Shortly afterwards supernovae were proposed as a comparable
source of refractory dust \citep{Hoylewick70}.

Both direct observations and theory have now converged to leave little
doubt that mass outflows from asymptotic giant branch (AGB) stars,
including carbon stars, as well as supernovae, are sources of
refractory dust in interstellar space.  However, by analysing the
quantity of dust in the Galaxy as well as in external galaxies (SMC,
LMC, Andromeda satellites), it is possible to set stringent limits on
the contributions from different stellar dust sources
\citep{Draine2009}.  \cite{Matsuura2009} investigated the global mass
budget in the LMC and identified a missing dust-mass problem. The
accumulated dust mass from AGB stars and SNe was estimated to be
significantly lower than observed in the ISM.  A similar conclusion was
reached by \citep{Zhu2015} who also included common
envelope ejecta and by \cite{Srinivasan2016} from studies of the
SMC.  \cite{DeLooze2016} investigated the origin of interstellar dust
in the Andromeda dwarf galaxies and concluded that the observed dust
content is an order of magnitude higher than expected from AGB stars and
SNe.  It now appears that less than 10\,per cent of the dust in all
cases can be attributed to AGB stars and supernovae, thus requiring
the bulk of interstellar dust to form by grain mantle growth in the
denser clouds.  With a wide range of local conditions, cloud densities
in particular, a uniform grain composition as well as size is then
difficult to explain.  It is also not clear how grains with large
organic or inorganic mantles can be effectively dispersed into the
diffuse ISM where the extinction of light from distant stars ($d
> 1\,${kpc}) is observed.  In section~3 we shall argue that
biologically generated particles could account for the bulk of the
interstellar grains which possess a more or less invariant composition
and size.

The contribution to interstellar grains from the biological model
(discussed in section 2.5), namely biological dust, can be estimated
by noting that the present Oort cloud in our solar system, comprised
of $10^{12}$~comets, is generally assumed to have a loss rate to
interstellar space of about $10^{11}$~comets every $10^{9}\,$yr
\citep{Weissman1983}.  This is equivalent to the dispersal of
bacterial dust from about 100~comets every year.  Assuming a 10\,per
cent mass fraction of bacterial cells in comets of typical radius of
about 10\,km, the rate of injection of bacterial grains from our solar
system into the interstellar medium is $10^{19}\,\rm g\,yr^{-1}$.
With a total Galactic population of $10^{11}$~(G dwarf) solar-like
stars each endowed with a similar Oort cloud of comets, the total rate
of replenishment of bacterial dust into the ISM is $10^{31}\,\rm
g\,yr^{-1}$.  Distributed throughout the volume of the ISM,
about $10^{66}\,\rm cm^{3}$, this gives a rate of increase of grain density of
$10^{-35}\,\rm g\,cm^{-3}\,yr^{-1}$.  Over a typical turn-over time
of the interstellar medium, $3\times 10^{9}\,$yr, this leads to a mean
density of $3\times10^{-26}\,\rm g\,cm^{-3}$, which is essentially
most of the interstellar dust.  Notwithstanding the approximate nature
of these estimates, it would appear that the postulated bacterial
grains could make a significant, perhaps even dominant, contribution
to solid particles in the diffuse interstellar medium.

\subsection{Constraints from Interstellar Extinction}

The visual extinction of starlight in the solar vicinity,
predominantly arising from scattering rather than absorption, amounts
to about $1.8\rm\,mag\,kpc^{-1}$ \citep{Lyndwick68}.  This demands
a total mass density of grains in the form of silicates, ices and
carbonaceous/organic grains with optimum radii to produce visual
extinction (mainly owing to scattering) of approximately
$0.3\,\mu$m. This material accounts for over 30\,per cent of the
available C, O and Si in the interstellar \citep{HW91}.  Grains
growing much larger mantles in the dense interstellar clouds are
excluded by this criterion.

The wavelength dependence of extinction has continued to provide
stringent constraints on the dust and for any particular model, such
as graphite--silicate mixtures, the particle size spectrum and optical
properties need to be finely tuned to agree with the more or less
invariant shape of the interstellar extinction curve
\citep{Draine2003}.

\begin{figure}
\begin{center}
\includegraphics[width=\columnwidth]{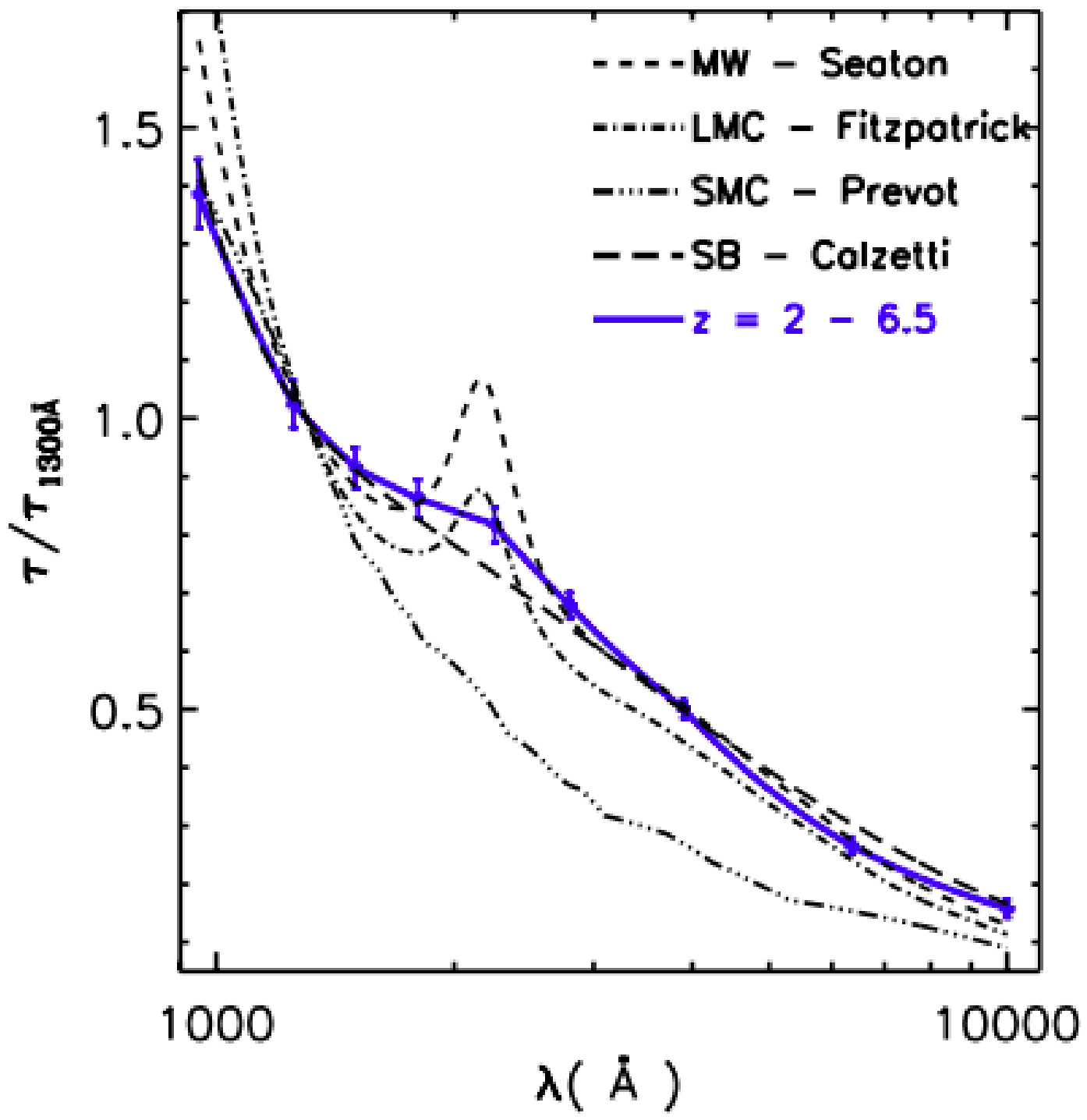}
\caption{Extinction curves for galaxies of various redshift.  Milky Way
  (MW) \citep{Seaton1979}, LMC \citep{Fitzpatrick1986}, SMC
  \citep{Prevot1984}, local starburst galaxies \citep{Calzetti2000}.
  Compilation is due \cite{Scoville2015}.}
\label{Fig1}
\end{center}
\end{figure}

This interstellar extinction curve (Figure~\ref{Fig1}) has been
extended both with respect to the wavelengths covered and the
distances of sources examined.  For wavelengths long-ward of
3000\,{\AA} it is known to be more or less invariant compared to that
for shorter ultraviolet wavelengths \citep{HW91}. The main variation
in the ultraviolet is in the slope and the extent of a broad symmetric
hump in extinction always centred at 2175\,\AA.

Extinction curves for the LMC and the SMC have been observed since
1980 \citep{Gordon2003}.  That for the LMC is found to be more or less
identical to the Galactic extinction curve while that for the SMC is
different to the extent that it possesses a weak or non-existent
2175\,{\AA} bump.  Extinction curves have also recently been extended
to very distant galaxies \citep {Scoville2015, Noll2009}.

It has been widely assumed that the most plausible explanation for the
2175\,{\AA} feature of interstellar dust involves a graphite particle
model \citep{HW62}.  \cite{WickGuil65} first used laboratory optical
constants of graphite, together with the Mie formulae
\citep{vanderhulst1957}, to compute the extinction cross-sections of
graphite spheres of various radii.  All subsequent calculations for a
graphite grain model have followed basically the same procedure.  An
unsatisfactory feature of the graphite particle explanation of the
2175\,{\AA} absorption is the requirement that the graphite is in the
form of spheres with a finely tuned radius of $0.02\,\mu$m and,
moreover, that it has isotropic optical properties.  The latter
assumption is not physically realistic because graphite has strongly
anisotropic electrical properties. The particular choice of size in
all current modelling of this feature is difficult to defend
\citep{HW91} and also appears inconsistent with recent theoretical
predictions of grain sizes from stellar outflows of AGB stars of solar
metallicity \citep{Del2017}.

A suitable model for the 2175\,{\AA} feature must explain (1)~the
observed invariance of the rest central wavelength independent of red
shift (at least up to $z \approx 6$) and (2)~the evidence that the
strength of the feature depends on galactic environment, with more
active galaxies showing weaker features (Figure~\ref{Fig1}).  These
requirements are fulfilled by a dispersed macromolecular organic
absorber of universal prevalence.  In more active galaxies it would be
reasonable to assume that the stability of such molecules is
compromised to varying degrees.  Hence variations of the strength of
the 2175\,{\AA} feature would be explained.

Polyaromatic hydrocarbons (PAHs) have been identified as possible
candidates to explain the invariant wavelength of the 2175\,{\AA}
ultraviolet absorption feature.  Owing to variability of the central
wavelength of the band for different PAHs, for such an explanation to
be viable, a highly specific distribution of PAHs with varying states
of ionization is demanded \citep{HW2000,Lidraine2001}.
  
We propose that the universality of this feature can be plausibly
explained if the macromolecular absorbers are of biological origin.
Aromatic and hetero-aromatic groups, $\rm C_6$ rings and rings with O
or~N inclusions, form a large class of organic molecules derived from
biology and may represent the most stable break-down products of
living systems.  Such molecules including quinolone and quinozoline
are found to have a broad absorption feature resembling the
2175\,{\AA} interstellar absorption feature \citep{HW1977, HW1979}.  It
would appear that this feature is best matched by the subset of
aromatic functional groups found in biology.

\subsection {Abiotic explanations of IR emission features}

The unidentified infrared bands (UIB's) at 3.3, 6.2, 7.7, 8.6
and~$11.3\,\mu$m seen in the diffuse ISM arising from CH and~CC
vibrational modes of aromatic compounds are usually attributed to
PAHs.  However this association is by no means universally
accepted.  When UIBs are seen in astronomical sources they are usually
also associated with aliphatic bands at 3.4, 6.9 and~$7.3\,\mu$m, so
the basic PAH model has been modified to include methyl side groups.
As an offshoot of such models, \cite{Zhang2015b} have proposed a
class of mixed organic aromatic/aliphatic nanoparticles (MOANs) as a
possible carrier of the UIB bands.  Importantly, they also demonstrate
that a general agreement between astronomical spectra and spectra of
PAH mixtures, as is often found, is not proof of the PAH hypothesis.

As we pointed out earlier, a case can be made for the presence of
fullerenes in the ISM, but again the question arises as to how they
are formed.  Their discovery in a proto-planetary nucleus suggests
formation on a time-scale of $10^3$\,yr which presumably excludes a
bottom-up process because of the time required to build
macro-molecules from smaller carbon units \citep{berne2015}.  The favoured mechanisms
are photon or shock induced processes starting from large hydrocarbon
molecules leading to de-hydrogenisation and ejection of $\rm {C_2}$
\citep{Micelotta2012,omont2016}.  Possible precursors are PAHs, HACs
(hydrogenated amorphous carbon), and~MOANs.  Biology also produces
possible pathways for the generation of fullerenes.

\subsection {Organic molecules in proto-planetary discs}

We expect some fraction of the complex organic solids seen in the
diffuse ISM to have originated in proto-planetary nebulae.  In addition
to the ices that form in situ, dust particles in outflows from stars
are entrapped in the collapsing gas clouds, that form proto-planetary
discs, then reprocessed in various ways and ejected back into the
diffuse ISM.  In our alternative biological hypothesis, cometary
bodies, moons and planets that also form in these discs are the sites
where an incipient biology is amplified and then dispersed through a
galaxy.

Since the work of \cite{Sagan79} it has been known that, under
suitable conditions, intractable organic polymers are produced on
grain surfaces in laboratory experiments devised to simulate
conditions in proto-planetary discs \citep[see][for more recent work
  in this area]{Munoz2002}.  So the discovery that a large fraction of
interstellar carbon in the diffuse interstellar medium is tied up in
the form of PAHs and other organic molecules relevant to life has been
taken as evidence of such grain surface chemistry or ion--molecule
reactions occurring on the surfaces of grains or within interstellar
grains irradiated by low-energy electrons and X-rays
\citep{Ehrenfreund2000}.  These are considered as plausible
non-biological sources of the organic component of interstellar dust
in the diffuse ISM.  While it is true that the laboratory polymers
formed in this way possess the generic functional groups that are
present in all organic material and that similar processes may indeed
occur in different interstellar environments, a precise match to
astronomical observations must require arbitrary fixing of the
relative weights of such functional groups.  As we show in
section~3.1, the weightings that are required are those that occur
naturally through biology.  Furthermore, achieving densities of this
material exceeding $10^{-7}$ of the ambient hydrogen density in such
environments appears to present a serious challenge \citep{Walsh2014}.

There are also other difficulties associated with this class of model.
The feasibility of production of complex organics by the above
processes generally demands a reducing not an oxidizing mixture of
molecular ices dominated by $\rm {CH_4}$.  However, since the early
work of \cite{Oort1946}, the main component of interstellar ices has
been expected to be $\rm{H_2O}$ rather than $\rm{CH_4}$ and the
mixture of molecules in the condensed state would possess an overall
oxidizing composition.  This is indeed consistent with the observation
that the dominant form of C in the gaseous interstellar medium is
$\rm{CO}$ rather than $\rm{CH_4}$ or CN.  So the conditions under
which laboratory experiments have thus far been carried out, involving
molecules such as $\rm{CH_4}$ \citep[for example][]{Esmaili2017}, are inapplicable to the
bulk of interstellar ice grains or ice mantles.  They could only be
justified as being relevant to localised regions of interstellar
clouds.  Because thermodynamics must always prevail on the large
scale, it is unlikely, in our view, that these processes have a role
to play in the bulk chemical transformation of astrophysical ices
leading to biologically relevant compounds

\section{Case for Biological Grains}

A strong argument in favour of biology is that wherever and whenever
it operates it is incomparably more efficient than competing abiotic
processes.  This is apparent when we look at the Earth where
99.999\,per cent of all organics have a biological origin
\citep{HW2000}.  The question is to what extent such processes can
operate and convert inorganic matter to organic matter in the physical
conditions expected in different astronomical environments.  Here,
except in the cases of comets and solar system planets and moons, we
need to be guided by what we can deduce from remote observations of
the properties of interstellar matter.

\subsection {The 2.9 to 4$\,\mu$m absorption spectrum of GC-IRS7} 

\begin{figure}
\begin{center}
\includegraphics[width=\columnwidth]{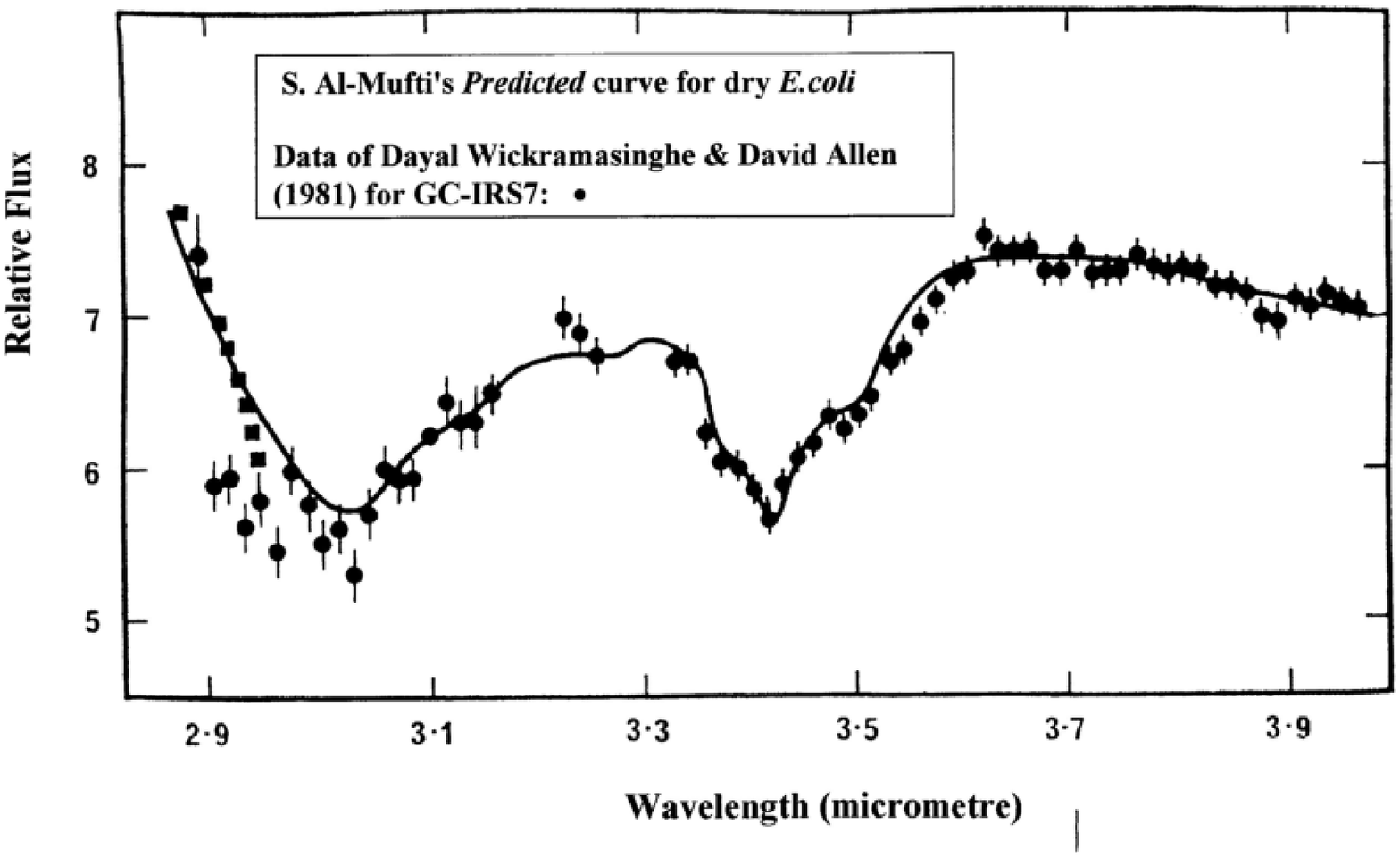}
\caption {The first detailed observations of the Galactic Centre
  infrared source GC-IRS7 \citep{Allen1981} over the spectral region
  2.9 to~$3.6\,\mu$m obtained with the AAT compared with
  laboratory spectral data for dehydrated bacteria \citep{Hoyle1982b}.
  The $3.4\,\mu$m feature corresponds to a mixture of aliphatic and
  aromatic CH stretching modes and the broad $2.9\,\mu$m feature to a
  combination of vibrational/rotational modes of OH bonds.}
\label{Fig2}
\end{center}
\end{figure}

In contrast to the extinction in the visual band which is mainly due
to scattering, the infrared and ultraviolet absorption features of
interstellar dust are independent of grain size and yield invaluable
information on its molecular composition.  The possibility of a
largely organic model of interstellar dust became apparent when the
first mid-infrared observations of the source GC-IRS7 near the
Galactic Centre in the spectral region 2.9 to~$4\,\mu$m with the
Anglo-Austrlian Telescope (AAT) revealed the average properties of
dust over a distance scale of some 10\,kpc \citep{Allen1981}.  The
absence of the usually dominant water ice feature at $3.06\,\mu$m in
observations towards most molecular cloud sources demonstrates that
the path length towards GC-IRS7 represents the properties of dust in
the diffuse ISM.  The discrete features stretching from 2.9
to~$3.6\,\mu$m can be identified with a mixture of aliphatic and
aromatic CH stretching modes in an organic solid combined with
vibrational/rotational modes of OH bonds.  \cite{Hoyle1982b} showed
that there is a close correspondence between the astronomical data and
laboratory data of desiccated bacteria shown in Figure~\ref{Fig2}.
The total extinction over a 10\,kpc path length to the source GC-IRS7
is 0.56\,mag.  Combining this data with the measured absorption
coefficient of desiccated bacteria at $3.4\,\mu$m of about $750\,\rm
cm^2\,g^{-1}$ we find that the smeared out density of absorbing
material in interstellar space is $2\times 10^{-26}\,\rm{g\,cm^{-2}}$
close to 50\,per cent of the density of dust responsible for the
visual extinction of starlight.

The $3.4\,\mu$m spectral feature in GC-IRS7 stands out in support for
the theory of cosmic biology.  This feature has now been confirmed by
observations of other Galactic Centre sources that do not show a
dominant water-ice feature.  We note in particular the study of the
Quintuplet region by \cite{Chiar2013} who, using the United Kingdom
Infrared Telescope (UKIRT), find spectra are similar to the GC-IRS7
spectrum.  The small variations can be interpreted as representing
varying states of degradation of bacterial grains expected in
different environments.

The conclusion is that the bulk of interstellar dust contains a
distribution of CH and~OH bonds, in various configurations, that
closely matches biology.  The details of such spectroscopic
correspondences cannot be claimed to be either unique or incompatible
with more conservative dust models (section~3.3).  However the
existence of such a high fraction of interstellar carbon in the form
of organic molecules linked together so as to mimic biology raises a
tricky problem relating to the origin of such material.  Biology itself
is one logical option but, if that is to be rejected, abiotic
synthesis is required and the questions of where this occurs and its
efficiency have to be answered.  The problems associated with one such
possibility, namely the synthesis of complex organic molecules in
proto-planetary discs, have already been discussed (section~2.5).

\subsection {Extended Red Emission}

\begin{figure}
\begin{center}
\includegraphics[width=\columnwidth]{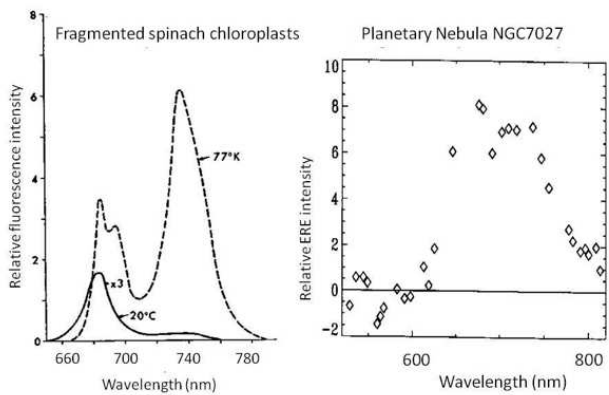}
\caption {Left panel: Spectra of fragmented spinach chloroplasts at
  two temperatures \citep{Boardman66}.  Right panel: A spectrum of ERE
  excess in NGC7027.}
 \label{Fig3}
\end{center}
\end{figure}

The detection of extended red emission (ERE) over the waveband 6000 to
8000\,{\AA} in planetary nebulae \citep {Witt1985,Witt1984} can also
be interpreted as evidence for the presence of aromatic molecules
which could be of biological origin.  This feature has now been
ubiquitously observed in a wide variety of dusty objects and regions
in our Galaxy and outside.  ERE has also been observed in the diffuse
interstellar medium \citep{Gordon98} and in high latitude Galactic
cirrus clouds \citep{Witt2008}.  \cite{Witt2004} have reviewed
abiotic models to interpret ERE emission by a hypothetical
photoluminescence process in interstellar dust containing aromatic
groups.  The absorption of photons at ultraviolet/optical wavelengths
is postulated to be followed by electronic transitions associated with
the emission of longer-wavelength optical and near-IR photons as a
two-stage de-excitation process near the valence band in a
semiconductor particle.  Although it is widely claimed that ERE is
caused by simple, possibly compact, PAHs under various excitation
conditions, the actual fits of models to the astronomical data have
not been satisfactory \citep{Thaddeus2006}.  The fluorescence in
fragments of biomaterial such as chloroplasts, cell inclusions
including chlorophyll, offers another possibility.  Figure~\ref{Fig3}
compares the fluorescence behaviour of fragmented spinach chloroplasts
\citep {Boardman66} at different temperatures with the observations of
a planetary nebula NGC7027 \citep{Furtonwitt92}.  We note the presence
of two temperature dependent broad features centred at about
6800\,{\AA} and 7400\,{\AA} which may make a contribution to the
observed excess in NGC7027.  Chloroplasts are not presented here as a
definitive or unique identification of the ERE carrier but rather as
an illustration of the type of biological PAH that could collectively
play a role.  It is also worth noting that the biological structures
that give rise to ERE could also be responsible for many of the
observed UIBs \citep {Smith07, Kwok09, Kwok2011}.

\subsection {Biotic or abiotic origin?} \label {subsec:biol}
  
The question now arises as to the mode of origin of interstellar
organic material that appears similar to desiccated biomaterial and
its component products.  We have already mentioned that the preferred
formation mechanisms, that involve grain surface chemistry, grain
mantle processing and ion-molecule reactions, although technically
feasible, have uncertain efficiencies.  The convergence of all such
processes to produce a desired end result in terms of satisfactory
spectral fits to astronomical data sets is also arbitrary.  In one
particular instance, the broad $3.4\,\mu$m absorption profile of the
Galactic Centre Quintuplet sources (section~3.1) has
been shown to be reproducible with a carefully chosen laboratory
mixture of compounds with appropriate aliphatic and aromatic CH
stretching modes \citep{Chiar2013}.  Whilst the fits to the data
cannot be disputed, its applicability to only a limited wavelength
region and the ad hoc nature of the weighting factors involved in the
synthetic spectrum has no physical basis.

Universal biology on the other hand would set the relative proportions
of its main functional chemical units, carbohydrates, lipids, nucleic
acids, proteins, within narrow limits and their eventual degradation
to coal and graphite would also be expected to follow a natural course
\citep{McFadzean1989,Okuda1990,Ishii1998}.  Mixtures of biomaterial and
inorganic dust from AGB stars and supernovae could in our view offer a
plausible explanation of all the available astronomical data.

\section{Comets as a source of interstellar microbiota}

The view that comets and similar icy bodies are unsustainable as
microbial habitats is based on the premise that their surface
temperatures at heliocentric distances very much greater than 1\,AU
lead inevitably to hard-frozen conditions.  This is not true if
radioactive nuclides are taken into account \citep
{HW78,Wallis1980}.  \cite{Yabushita1993} has shown that with the
inclusion of radioactive nuclides $\rm{^{232}Th}$, $\rm{^{238}U}$ and
$\rm{^{40}K}$, liquid water domains in comets can be maintained for
more than a Gyr after they first condense in the outer regions of the
solar nebula.  Within such domains, even very small numbers of viable
bacterial cells or spores could grow exponentially on a very short
time-scale.  Biological activity occurring in primordial comets at
heliocentric distances greater than 20,000\,AU can then contribute to
ejecta that reach escape velocities from the entire solar
system.  In our own solar system the total mass of comets and
icy planets is about $10^{-2}$\,\Msun so we can argue generally that a
significant fraction of the C, N and~O material that goes into star
formation is processed into bacterial type-dust within cometary bodies
and ejected back into interstellar space.  Within the uncertainties of
all the assumptions involved in this argument we do have a process
that can produce dust at higher rates than supernovae and AGB stars.
Furthermore, in comets approaching perihelion, biological activity
could be transiently triggered by solar heating and so lead to the
generation of high-pressure pockets of gaseous metabolites that in
turn trigger the release of jets of gas and biological particles in
their wake \citep{Wick1996}.  Similar conditions are likely to exist
in subsurface lakes in other solar-system planetary bodies including
Europa and Enceladus.

\subsection {Biological material in comets}

When \cite{HW81} first promulgated the idea of cometary panspermia,
the supporting evidence was tenuous and appeared far-fetched.  There
was early evidence from infrared spectroscopy that organic polymeric
grains were present in the dust tails of comets, and thus called into
question the Whipple dirty snowball paradigm \citep {Vanysek1975}.
There is also the observation that cometary nuclei can be coal-like
with low albedos, in contrast to the expected high albedos of icy
surfaces \citep{cruikshank1985}.

\begin{figure}
\begin{center}
\includegraphics[width=\columnwidth]{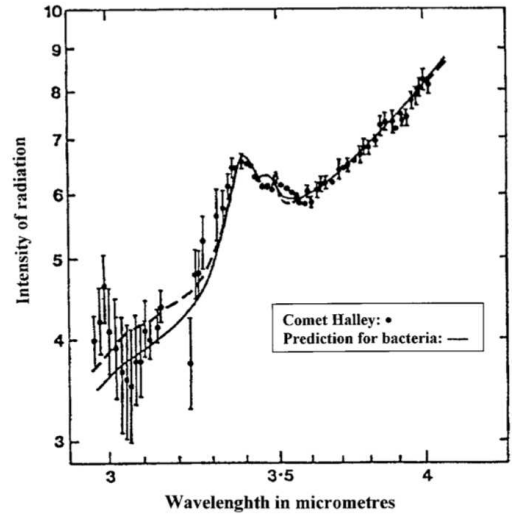}
\caption {Emission by the dust coma of Comet Halley observed on March
  31, 1986 (points) using the AAT compared with normalised fluxes for
  desiccated E-coli at an emission temperature of $320\,$K.  The solid
  curve is for unirradiated bacteria; the dashed curve is for X-ray
  irradiated bacteria \citep {WickAllen1986, HW2000} }
 \label{Fig4}
\end{center}
\end{figure}

The idea of cometary microbiology moved swiftly from speculation to
quantitative measurement after the last perihelion passage of
Comet~P/Halley in 1986.  The first investigation of a comet in the
Space Age ({\it Giotto} mission) marked an important turning point in
the history of cometary science.  A dark comet surface, darker than
the darkest coal, was observed by the Giotto photometry.  More
importantly, \cite{WickAllen1986}, using the Anglo-Australian
Telescope, obtained a 2 to~4$\,\mu$m spectrum of the dust from an
outburst of the comet on 31st March 1986.  This spectrum showed
unequivocal evidence of CH rotational/vibrational stretching modes
indicating complex aromatic and aliphatic hydrocarbon structures,
consistent with the expected spectrum of bacterial dust
\citep[][Figure~\ref{Fig4}]{HW2000}.

A number of similar discoveries soon followed.  The infrared spectrum
of Comet Hale--Bopp at 2.9\,AU \citep{Crovisier1997} was shown to match
the behaviour of a mixture of microbes with a 10\,per cent
contribution of silicates predominantly contributing only to the
$10\,\mu$m feature \citep{WH1999}.  A very similar spectrum was
obtained from the post-impact ejecta in the {\it Deep Impact} mission of
2005 \citep{Ahearn2005,Lisse2006}.  \cite {Wick1996} had argued that
the prodigious output of organic dust and CO observed from Comet
Hale--Bopp at 6.5\,AU, where no cometary activity was normally
expected, was indicative of the high-pressure release of material from
a liquefied subsurface domain within which microbiology may exist.

Historically the next significant correspondence with biology was to
emerge from the {\it Stardust} mission which captured high speed
cometary dust in blocks of aerogel and studied the residues in the
laboratory.  Amongst these was the most common of amino acids glycine
together with a complex mixture of hydrocarbons \citep{Elsila2009}.

The conclusions that can be drawn from these and similar studies,
based on space missions to comets, is that the evidence so far does
not contradict cometary microbiology.  Rather, there are unexpected
features that continue to be discovered with the most recent space
probes of comets, that are inconsistent with expectations of simple
chemistry but are readily explained in terms of cometary biology.

\subsection{Thermochemical equilibrium or biology - new evidence?}

Whenever a distant astronomical body exhibits conditions that can
support life, the next objective is to look for evidence of molecules
or chemistry that may be indicative of life.  The model for life that
is used is of terrestrial carbon-based life, the only life we know of
and available for direct experimentation.

\begin{figure}
\begin{center}
\includegraphics[width=\columnwidth]{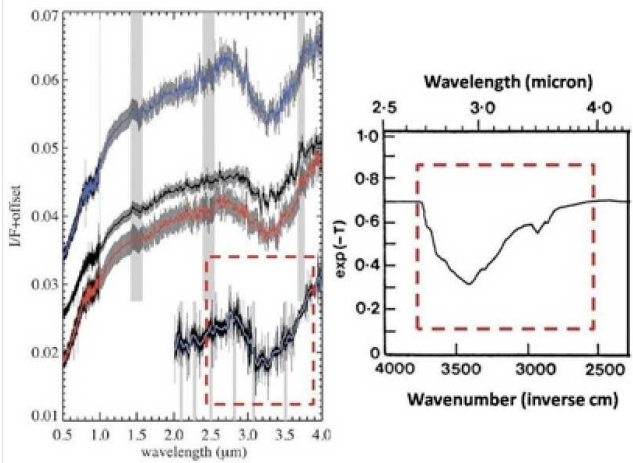}
\caption{Left panel: The surface reflectivity spectra of
  comet~67P/C--G \citep{Capaccioni2015} taken at different times.
  Right panel: The transmittance spectrum of desiccated E-coli
  (cf. Figures \ref{Fig2} and~\ref{Fig4}) over approximately the same
  wavelength range for comparison.}
\label{Fig5}
\end{center}
\end{figure}

The {\it Rosetta} mission's {\it Philae} lander has recently provided
information about the comet~67P/C--G that appears to be in conflict
with the concept of a hard-frozen non-biological comet.  The overall
reflectivity spectrum of the surface material of the comet is similar
to that of biological material as seen in Figure~\ref{Fig5}
\citep{Capaccioni2015,Wick2015}.  Jets of water and organics issue
from ruptures and vents in the comet's frozen surface could be
consistent with biological activity occurring within sub-surface
liquid pools.

A recent report of $\rm{O_2}$ along with evidence for the occurrence
of water and organics together in the same place \citep{Bieler2015}
provides further indication of ongoing biological activity.  Such a
mixture of gases cannot be produced under equilibrium thermodynamic
conditions because organics are readily destroyed in an oxidizing
environment.  The freezing of an initial mixture of compounds,
including $\rm{O_2}$, not in thermochemical equilibrium has been
proposed. Although possible, there is no evidence to support such a
claim.  On the other hand the oxygen/water/organic outflow from the
comet can be readily explained on the basis of subsurface
microbiology.  Photosynthetic microorganisms operating at the low
light levels near the surface at perihelion could produce $\rm{O_2}$
along with organics. The combination of oxygen $\rm{O_2}$, $\rm{O_3}$
(oxidizing) with gases such as methane (reducing) is not permitted in
local thermodynamic equilibrium (LTE) so their coincidence in space is
possible evidence of biology.

We next turn to the presence of the amino acid glycine and an
abundance of phosphorus in the coma of comet~67P/C--G
\citep{Altwegg2016}.  The very high ratio of $\rm{P/C \approx
  10^{-2}}$ by number observed is difficult to reconcile on the basis
of volatilization of condensed material of solar composition with
$\rm{P/C\approx10^{-3}}$, particularly when we might expect inorganic
phosphorus to be mostly fixed in refractory minerals.  On the other
hand the $\rm{P/C}$ ratio of organic material found from {\it Rosina}
can be explained if the material in the coma began as biomaterial, such as
virions or bacteria.

Methyl chloride ($\rm{CH_3Cl}$), together with other haloalkanes, is a
significant source of atmospheric carbon via volatile haloalkanes on
Earth.  It is mostly produced by oceanic biochemical processes by
various macroalgal and other biomasses both in the presence and
absence of sunlight \citep {Lobert1999,Lat1998,Lat2001}.  Further
emission of $\rm{CH_3Cl}$ by wood rotting fungi contributes to some of
the atmospheric complement of this molecule \citep{Watling1998}.  So
the presence of significant amounts of $\rm{CH_3Cl}$ and related
haloalkanes in space is a possible biomarker, indicating production in
habitats similar to those found on Earth.  The discovery of this
molecule both in interstellar clouds and in comet~67P/C--G
\citep{Fayolle2017} is therefore significant.

Finally we note the intriguing observation that comet Lovejoy emits a
prodigious amount of ethanol, equivalent to that in 500\,bottles of
wine, per second \citep{Biver2015}.  This has no explanation in terms
of simple chemistry.  On the other hand many species of fermenting
bacteria can produce ethanol from sugars so this may be another
indication that microbial processes operate in comets.

\subsection {Space hardiness of microbes}

The idea that microbial life does not exist in comets still dominates
scientific orthodoxy but its rational basis is fast eroding.  There is
ample evidence that microorganisms are space hardy in every
conceivable respect \citep {Wick2015b,Horneck2001}.  There are many
types of cryophilic microorganisms that are resistant to freeze--thaw
cycles and it has recently been demonstrated that such microbes exist
in arctic ice and permafrost.  They can even metabolise at temperatures
well below the freezing point of water \citep{Junge2006}.

Early objections to the concept of panspermia were based on the
expectation that bacteria and other microorganisms would not survive
space travel, high speed entry into a planetary atmosphere or long
periods in space \citep{Horneck2001}.  However, laboratory experiments
have shown that microorganisms can survive hypervelocity impacts
\citep{Burchell2004a}.  Microorganisms with even modest protective
mantles against UV radiation, and it seems likely that microorganisms
encased in meteorites, can survive in interstellar space for millions
of years \citep{Horneck1993,Horneck1994,Horneck1995}.  Microbes have also
been shown to survive on the exterior of the surfaces of rockets fired
through the atmosphere \citep{Thiel2014}.

The search for incoming bacteria and viruses with balloon-borne
equipment lofted to the stratosphere has been carried out for nearly
two decades \citep{Harris2002,Wainwright2003,Shivaji2009}.
Stratospheric air samples recovered from heights ranging from 28
to~41\,km yield evidence of microorganisms, but these are widely
regarded as most likely to have been lofted from the Earth's surface.
Recently \cite{Grebennikova2018} have confirmed the discovery of
several microbial species associated with dust on the exterior windows
of the {\it International Space Station} at a height of 400\,km above
the Earth's surface and contamination at source and in the laboratory
has been ruled out.  The results of PCR (polymerase chain reaction)
amplification followed by DNA sequencing and phylogenetic analysis
have established the presence bacteria of the genus Mycobacteria and
the extremophile genus Delftia, amongst others, associated with
deposits of dust. \cite {Grebennikova2018} discuss possible mechanisms
for ejecting terrestrial microbes against gravity to such heights in
the ionosphere but concede that an origin from outer space remains a
possibility.

Far from being the fragile earth-oriented structures they were once
considered, microorganisms are incredibly space-hardy. In a review of
these and other similar experiments, \cite{Burchell2004b} concludes
that, with the data at hand, interstellar panspermia can neither be
proved nor disproved.  It therefore remains a logical option to be
evaluated on the basis of other data as we discuss here.

In order for panspermia to operate only a minute fraction of microbes
need to survive the transit between one amplification site, such as
comets in one planetary system, to another.  The exponential
replication power of microorganisms is well recognised much to the
chagrin of hapless victims of microbial diseases on Earth!

\subsection {Exo-planets and Exo-comets}

Recent observations from the {\it Kepler} mission have shown that
planetary systems are exceedingly common in the Galaxy.  The total
tally of habitable exoplanets is estimated as upwards of~$10^{10}$
\citep{Kopparapu2013} and, because comets and asteroids form an
integral component of our solar system, it is reasonable to presume
that such objects are also common in other planetary systems.  The
detection of exo-comets is not easy, and only in one case, around the star
KIC2542116, is there direct evidence of transiting comets \citep{Rappaport2018}.

A significant fraction, perhaps 30\,per cent, of long-period comets
that approach perihelion actually leave the solar system in hyperbolic
orbits \citep{hughes1991}.  If other exo-planetary systems are
similar, and are losing comets at a comparable rate, such comets must
be reaching us at a steady rate.  The first confirmed interstellar
object in our vicinity Oumuamua \citep{Meech2017}, may be such an
object if it is found to be an extremely low albedo comet rather than
an asteroid \citep{Wickramasinghe2018}.  There are also indications
that for the star KIC8462852 the unusual dips in brightness (up to
20\,per cent over timescale of days, as well as longer period
oscillations that continued to the present day, may be due to
obscuration by dust from comets \citep{Wyatt2018}.  Further detection
of exo-comets will provide necessary evidence for or against the
possibility of interstellar panspermia.

\section {Comets and Interstellar dust - a re-appraisal}

The data from comets corroborates the hypothesis that similar data
relating to the organic composition of interstellar dust, starting
with the observations of \cite{WickAllen1980} and \cite{Allen1981} and
subsequent observations of interstellar dust and molecules in
different astronomical environments \citep[see][for a recent
  review]{Kwok2016}, may have a biological origin. Indeed it has become
increasingly apparent in all these cases that it is not possible to
exclude biology with remote spectroscopy alone.  The spectroscopic
signatures of biology or the detritus of biology are, to the first
order, very similar to what is seen in astronomy.

It has been claimed that the data currently at hand already contradict
the hypothesis of cosmic biology.  We now examine these claims one by one.
\begin {itemize} 
\item
One of the key quantities that can be estimated from unmanned space
probes of comets is the average composition of outflows.  It might be
claimed that the near solar-system type compositions that come out
from such studies rule out the possibility of biology in comets
\citep{Kwok2016}.  However, the determination of the {\it average}
chemical composition of an astronomical environment such as a comet,
that must necessarily contain only a small fraction of biological
material, cannot provide a discriminant that would either rule in or
rule out biology.  Other considerations such as we have discussed need
to be properly assessed and evaluated.

\item
A specific objection raised against biology in comets is that no
molecules exclusive to biology, such as ATP or DNA, have been
discovered.  This objection appears to be particularly cogent for the
case of comet67P/C--G where some in situ spectroscopy has been carried
out.  However, in the absence of any direct life detection experiment
on the {\it Philae} lander we have to rely on indirect indicators that
are already strong.  These include the infrared reflectivity spectra
Figure~\ref{Fig5} over the 2 to~$4\,\mu$m region, the phosphorous
excess and the non equilibrium $\rm O_2$ abundances.  Moreover it
should be stressed that biological replication is supposed, according
to our model, to occur at some depth below the surface so that surface
organics may be mostly comprised of biological metabolites with little
in the way of living cells with detectable DNA.

\item
The match of the interstellar absorption and extinction data to
biology is not unique and the same data can be matched with arbitrary
mixtures of silicates, graphite and abiotic organic molecules
\citep{Draine2003}.  However, as we have already noted, in the case of
a non-biological explanation, ad hoc assumptions of relative
proportions need to be made and the precise invariance of these
proportions in different directions in the sky and in various
different astronomical and cosmological settings is difficult to
justify.  As we have discussed, universal biology leads to uniquely
defined proportions.

\item
A prevailing view is that the astronomical explanation is redundant
because biologists have a satisfactory and well-proven explanation of
abiogenesis on the Earth perhaps in deep sea thermal vents
\citep{Sojo2016}.  This is not true.  All attempts at re-creating
abiogenesis in the laboratory have failed.  No experimental proof of
abiogenesis in a terrestrial setting yet exists \citep{Deamer2012}, so
alternatives must be considered.

\end {itemize}

With the evidence currently at hand, and we confine the discussion to
astronomical data alone, the most we can rigorously claim is
consistency with a theory of biological grains.  However the single
unifying concept of cosmic biology remains perhaps the most economical
hypothesis to explain the array of facts available to us now.  The
facts to be combined derive from many diverse disciplines of
astronomy, biology, geology and space exploration.  An application of
Occam's razor prefers such a model.  Alternative explanations demand a
convolution of physical processes and mechanisms of varying degrees of
plausibility.

\section {Some predictions of Interstellar Panspermia}

Microorganisms on Earth display an incredible range of functional
diversity.  Particular microbial species colonise particular niches,
tolerating extremes of temperature, acidity, salinity, desiccation and
high levels of ionizing radiation.  Microbial habitats include
subterranean ocean caves, hydrothermal vents, polar ice and
tropospheric clouds.  Throughout this vast range of conditions an
amazing biochemical unity prevails pointing unerringly to a single
origin.  According to interstellar panspermia that event was probably
external to our solar system.

If panspermic life is discovered on planetary habitats outside Earth
such life is expected to possess the same unity of biochemistry that
is found on Earth, the same genetic code, the same DNA and similar
biochemistries.  Only the habitat selects what is best suited for
survival in a similar way to the microorganisms inhabiting the diverse
range of habitats on Earth.  If independent origin events occur such a
correspondence is extremely unlikely.  A radically different
biochemistry discovered elsewhere would be significant evidence
against a universal Galactic panspermia.

In the concordance Big Bang model of the Universe, our observable
horizon is estimated to have a radius of about 46\,G\,lyr
\citep{Tamara2004}.  Micro-wave background radiation observations
point to the Universe being flat to within 0.4\,per cent
\citep{Hinshaw2013}.  The true spatial extent of the Universe may even
be infinite if the it is precisely flat.  If life originates as a
highly improbable event, it is possible that such events are localized
within disjoint life domains in the Universe.  The cosmological domain
in which we happen to find ourselves defines the nature of our
idiosyncratic cosmic biology, including its basic biochemistry, as
well as its full range of genetic diversity, a range that is only
incompletely expressed on Earth.  Panspermia reduces the status of the
Earth to one of many building sites upon which an already evolved
cosmological legacy of life came to be expressed.

History of science has repeatedly shown any theory that requires the
Earth to hold a special position in the Universe has proved wrong, the
Copernican Principle.  The Earth is not the centre of our Universe, nor
is our Solar system nor our Milky Way Galaxy.  It is therefore
extremely unlikely that the Earth can in any way be regarded the
centre of life of the Universe.  This is particularly so if we take
account of recent discoveries of habitable exoplanets with an
estimated total of greater than $10^{10}$ in our galaxy alone.

Our case for panspermia rests on a few simple ideas.  First the very
early but punctuated origin of life on Earth seems difficult to
explain with a slow mutation and selection alone.  Secondly biological
molecules produced by terrestrial life provide a simple explanation
for what is observed in our Solar System, our Galaxy and further
reaches of the Universe.  Such uniformity is only possible if life
originated from a single or a number of very similar specialised
environments. That it has spread from such a single source provides a
simple and easily falsified explanation of a number of otherwise
difficult to explain observations.

We conclude by reminding the reader of the prevailing state of science
before the Ptolemaic solar system was finally abandoned.  Historian of
science Thomas Kuhn (1962) wrote,

{\it \textquotedblleft The state of Ptolemaic (Earth-centred)
  astronomy was a scandal before Copernicus' announcement.  Given a
  particular discrepancy, astronomers were invariably able to
  eliminate it by making some particular adjustment in Ptolemy's
  system of compounded circles.  But as time went on, a man looking at
  the net result of the normal research effort of many astronomers
  could observe that astronomy's complexity was increasing far more
  rapidly than its accuracy and that a discrepancy corrected in one
  place was likely to show up in another\textquotedblright}

\begin{acknowledgements}

  CAT thanks Churchill College for his fellowship.  Part of this work was
completed when CAT was visiting the Mathematical Sciences Institute of
the Australian National University and Monash University as a Kevin
Westfold distinguished visitor.

\end{acknowledgements}


\end{document}